\documentclass{ws-ijbc}

\usepackage{color} 

\usepackage[normalem]{ulem}
\usepackage{cancel}

\makeatother
\begin{document}
\catchline{}{}{}{}{} 

\markboth{A. Panchuk et al.}{Synchronization of coupled neural oscillators with heterogeneous delays}
\graphicspath{{figs/}}

\title{Synchronization of coupled neural oscillators with heterogeneous delays}

\author{Anastasiia Panchuk}
\address{Institute of Mathematics, National Academy of Sciences of Ukraine, Kyiv, Ukraine}
\author{David P. Rosin}
\address{Institut f{\"u}r Theoretische Physik, Technische Universit\"at Berlin, Hardenbergstra{\ss}e~36, 10623 Berlin, Germany}
\author{Philipp H\"{o}vel}
\address{Institut f{\"u}r Theoretische Physik, Technische Universit\"at Berlin, Hardenbergstra{\ss}e~36, 10623 Berlin, Germany\\
Bernstein Center for Computational Neuroscience, Humboldt-Universit\"at zu Berlin, Philippstra{\ss}e~13,
10115 Berlin, Germany}
\author{Eckehard Sch\"{o}ll}
\address{Institut f{\"u}r Theoretische Physik, Technische Universit\"at Berlin, Hardenbergstra{\ss}e~36, 10623 Berlin, Germany, schoell@physik.tu-berlin.de}

\maketitle

\begin{history}
\received{(to be inserted by publisher)}
\end{history}

\begin{abstract}
We investigate the effects of heterogeneous delays in the coupling of two excitable neural systems.
Depending upon the coupling strengths and the time delays in the mutual and self-coupling, the compound system exhibits different types of synchronized oscillations of variable period. We analyze this synchronization based on the interplay of the different time delays and support the numerical results by analytical findings.
In addition, we elaborate on bursting-like dynamics with two competing timescales on the basis of the autocorrelation function.
\end{abstract}



\section{Introduction}\label{sec:intro}
A plethora of synchronization phenomena has been found for coupled nonlinear oscillators in physical, chemical and
biological systems \cite{PIK01,BOC02,MOS02,BAL09,ZHA11}. Especially the interplay of synchronization and time delay in coupled systems has received much interest recently \cite{SCH07,JUS09,ATA10}. Delayed coupling plays also a crucial role in the case of oscillation death \cite{CHO07,ZOU09,ZOU12} and adaptive control schemes with delayed feedback applied to both chaotic \cite{WAN10a,LEH11a} and non-chaotic systems \cite{SEL12}. Previous studies involved networks of a large number of elements \cite{ATA04,DHA04,KIN09,CHO09,ZIG09,ENG10,BAT10,KAN11a,FLU11b} as well as simple recurring substructures consisting of a few systems only, so-called \textit{network motifs} \cite{HAU06,CHO07,DHU08,HOE09,FIE09,FLU09,BRA09,DHU11,HIC11,KYR11,ADH11}. Considering the dynamics on networks with delay, the local elements can be either time-continuous or time-discrete as for iterated maps \cite{WAN08c,WAN09c}. In addition, a number of universal model-independent results have been obtained \cite{FLU10b,HEI11}.

Synchronous dynamical patterns play also an important role in neuroscience 
\cite{ROS05,WAN05,HAU07a,MAS08,WAN08b,WAN09d,MAS09a,SEN09a,LIA09,LEH11,POP11}, where on the one hand they can be observed in 
ensembles of neurons as pathological states like migraine, Parkinson's disease, or epilepsy. On the other hand 
synchronization can also be beneficial for recognition, learning, or neural information processing. Obviously 
the signal transmission between neurons in different brain areas is not instantaneous. Thus non-zero transmission times 
have to be taken into account as crucial quantities that influence the dynamics of individual neurons to a large 
extent. Consequently effects due to time delays have attracted more and more attention in the studies of neural 
networks \cite{ROS05,HAU07a,MAS08,FRI09a,WAN10,WAN11c,LEH11,KAN11a,WAN11d} and particularly in motifs of two coupled neurons 
\cite{SCH08,DAH08c,HOE09,BRA09,HOE09a,HOE10}. The latter can be seen as the smallest entity in a larger network. 
Interestingly, phenomena observed in this area of research show a strong similarity with recent findings in 
optoelectronic oscillators \cite{ROS11a}.

Most previous works have assumed equal delay times in all connections. The focus of this paper, however, is on heterogeneous delays, which introduce additional timescales to the compound system. For this we consider a simple example of a network motif \cite{HAU06,DAH08c,PAN09,HOE09}, i.e., two delay-coupled neurons with delayed self-feedback, and we assume all delay times to be different. This configuration might as well be understood as two effective populations of larger clusters of neurons with delayed internal and mutual connections \cite{VIC08}.

The rest of this paper is organized as follows: Section~\ref{sec:model} introduces the neural model and the delay-coupling configurations. In Sec.~\ref{sec:identical}, we study interaction involving identical self-coupling delays numerically and analytically. The results are extended to the case of nonidentical self-coupling delays in Sec.~\ref{sec:nonidentical}. Section~\ref{sec:bursting} considers bursting dynamics. Finally, we close with a conclusion in Sec.~\ref{sec:conclusion}.

\section{Model}\label{sec:model}
\begin{figure}[th!]
  \centering
  \psfig{file=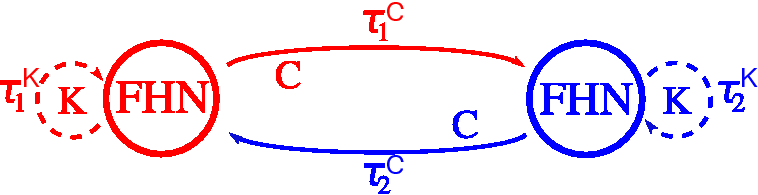,width=3in}
  \caption{Schematic diagram of two coupled neural elements including the parameters of the
mutual coupling (time delays $\tau_1^C, \tau_2^C$ and coupling strength $C$) and the self-feedback  (time delays
$\tau_1^K, \tau_2^K$ and feedback gain $K$).}
  \label{fig:fhn2model}
\end{figure}

We study a compound system of two coupled neural
elements each represented by a FitzHugh-Nagumo model
\cite{FIT61,NAG62}. The respective dynamic equations are paradigmatic for 
neural systems of type-II excitability, when periodic oscillations are generated in a Hopf bifurcation. We consider the
case where the two elements are
coupled such that each neural oscillator is subject to the delayed response from the other one. See
Fig.~\ref{fig:fhn2model} for a schematic diagram, where the time delays are denoted by $\tau_1^C$ and $\tau_2^C$ and $C$
is the coupling strength. In addition, we take also delayed self-feedback \cite{PYR92} with a delay time $\tau^K_i$, $i = 1, 2$ and feedback strength $K$ into account.

The two coupled FitzHugh-Nagumo systems are described by the following set of delay differential equations:
\begin{subequations} \label{eq:fhn2_dl_fdb}
\begin{eqnarray}
   \varepsilon_1\dot{x}_1 &=& x_1 - \frac{x_1^3}{3} - y_1 + C\left[x_2\left(t - \tau^C_2\right)- x_1(t)\right] \nonumber
\\
    &\,& + K\left[x_1\left(t - \tau_1^K\right) - x_1(t)\right] \\
    \dot{y}_1 &=& x_1 + a \\
   \varepsilon_2\dot{x}_2 &=& x_2 - \frac{x_2^3}{3} - y_2 + C\left[x_1\left(t - \tau^C_1\right)- x_2(t)\right] \nonumber
\\
    &\,& + K\left[x_2\left(t - \tau_2^K\right) - x_2(t)\right] \\
    \dot{y}_2 &=& x_2 + a,
\end{eqnarray}
\end{subequations}
where $\varepsilon_i$ denotes the timescale ratio between the slow inhibitor variable $y_i$ and the fast activator
variable $x_i$ ($i=1,2$). The parameter $a$ is known as threshold parameter. For $|a|>1$, the uncoupled system operates
in the excitable regime.  

As it was shown in Ref.~\citet{PAN09}, two coupled systems with asymmetric delay times $\tau_1^C$, $\tau_2^C$ can be reduced to a system
with symmetric delay times $\tau^C$. The difference between $\tau_1^C$ and $\tau_2^C$ leads only to a phase shift between oscillator 1 and 2. Therefore, we assume them to be equal $\tau_1^C = \tau_2^C = \tau^C$ without loss of generality and Eqs.~(\ref{eq:fhn2_dl_fdb}) can be
rewritten as follows
\begin{subequations}\label{eq:fhn2_dl_fdb_sym}
\begin{eqnarray}
   \varepsilon\dot{x}_1 &=& x_1 - \frac{x_1^3}{3} - y_1 + C\left[x_2\left(t - \tau^C\right)- x_1(t)\right]\nonumber \\
    &\,& + K\left[x_1\left(t - \tau_1^K\right) - x_1(t)\right]
    \label{eq:fhn2_dl_fdb_sym_x1}\\
    \dot{y}_1 &=& x_1 + a \\
   \varepsilon\dot{x}_2 &=& x_2 - \frac{x_2^3}{3} - y_2 + C\left[x_1\left(t - \tau^C\right) - x_2(t)\right] \nonumber \\
    &\,& + K\left[x_2\left(t - \tau_2^K\right) - x_2(t)\right] 
    \label{eq:fhn2_dl_fdb_sym_x2}\\
    \dot{y}_2 &=& x_2 + a.
\end{eqnarray}
\end{subequations}
Throughout this paper, we choose the following set of parameters, unless specified otherwise:
$\varepsilon_1 =\varepsilon_2 =\varepsilon = 0.01$, $a = 1.3$,
$\tau^C = 3$, and $C = 0.5$.

For $|a|>1$, the fixed point is always linearly stable (cf. Ref.~\citet{PAN09}) and the system Eq.~(\ref{eq:fhn2_dl_fdb_sym}) shows various regular spiking and bursting patterns. Furthermore, several stable
solutions can coexist for the same parameter values entailing high-level multi-stability. 

Before exploring the interplay between three different delay times, namely the mutual coupling delay $\tau^C$ and two nonidentical self-coupling delays $\tau_1^K
\neq \tau_2^K$ (Secs.~\ref{sec:nonidentical} and \ref{sec:bursting}), we will consider the case $\tau_1^K = \tau_2^K \equiv \tau^K$ in
the next Section. Analytical conditions for coherent spiking will be derived and generalized in the subsequent sections.

\section{Identical self-feedback delays}
\label{sec:identical}
The dynamics of the compound system~(\ref{eq:fhn2_dl_fdb_sym}) is diverse and hence, we will first introduce a classification. For this purpose, we will use an approach based on the mean interspike interval (ISI) $\left\langle T_j^{(i)}
\right\rangle \equiv T^{(i)}$, where $\left\{T_j^{(i)} \right\}_{j = 1}^N$ is the set of $N$ interspike intervals for
the time series of the $i$-th neuron ($i=1,2$) \cite{DAH08c,SCH08}. A measure based on the ISI is a powerful tool to
characterize regular, coherent spiking that is similar to a period-1 orbit. Therefore, we take the ISI values for those
cases into account that have small standard deviation. Failure of such a measure, e.g., for bursting dynamics, will be
discussed in later sections.

\begin{figure}[th!]
  \centering
  \psfig{file=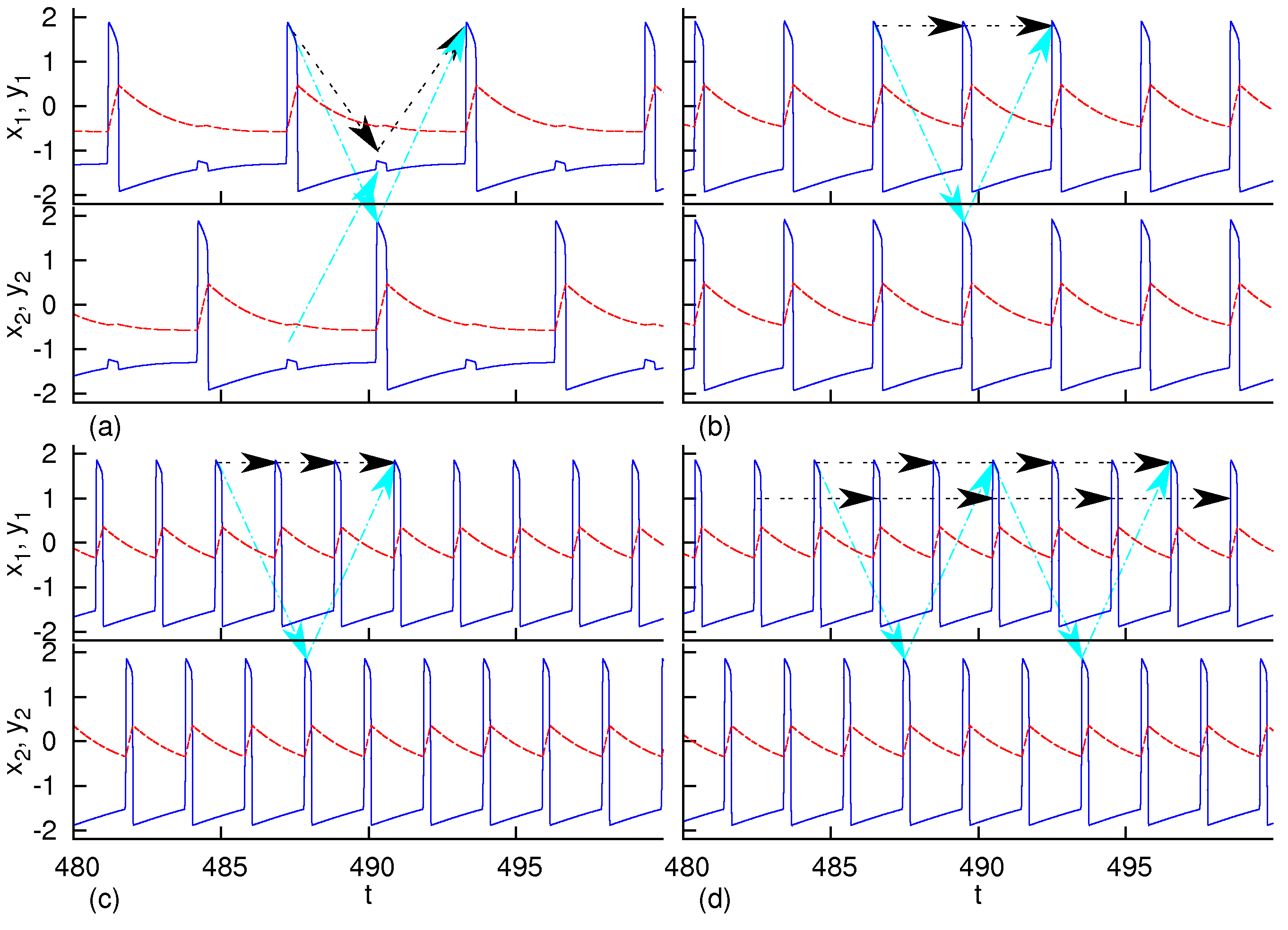}
  \caption{Time series of Eqs.~(\ref{eq:fhn2_dl_fdb_sym}) for identical self-feedback delays $\tau_1^K =
\tau_2^K = \tau^K$. The activator $x_i$ and inhibitor $y_i$ are shown by blue (full) and red (dashed) curves,
respectively. The parameters are chosen as (a) $K=0.05, \tau^K=3$, (b) $K=0.5, \tau^K=3$, (c) $K=0.5, \tau^K=2$, and
(d) $K=0.5, \tau^K=4$. The black and light blue arrows show excitations due to
self-feedback and mutual feedback, respectively. Other parameters: $\varepsilon = 0.01$, $a = 1.3$, $\tau^C=3$, and $C = 0.5$.}
\label{fig:samplespikes}
\end{figure}

Figure~\ref{fig:samplespikes} shows exemplary time series of coherent spiking for different values of $K$ and $\tau^K$,
while the mutual coupling delay and strengths are fixed at $\tau^C=3$ and $C=0.5$, respectively. The parameters are
chosen as ($K=0.05, \tau^K=3$), ($K=0.5, \tau^K=3$), ($K=0.5, \tau^K=2$), and ($K=0.5, \tau^K=4$) in panels (a)-(d),
respectively.
All these combinations of $\tau^K$ and $K$ exhibit coherent spiking, where the ISI is 
constant, hence the standard deviation vanishes. Since the fixed point in the individual subsystems is stable
in the excitable regime, we choose initial conditions such that only one subsystem is located in the fixed point
whereas the other is subjected to a one-time-only excitation. This initial excitation eventually remains in the
compound system due to the delayed coupling.

Comparing Figs.~\ref{fig:samplespikes}(a) and (b), one can see that small self-feedback gains, e.g., $K=0.05$ in
Fig.~\ref{fig:samplespikes}(a), are not able to trigger superthreshold excitations. Only subthreshold oscillations
occur after times $\tau^K$ as indicated by a black arrow in Figs.~\ref{fig:samplespikes}(a). In the following, we will
derive analytical conditions for the delay times $\tau^K$ and $\tau^C$ such that regular, superthreshold spiking occurs.

If the coupling strengths $K$ and $C$ are large enough, a spike at time $t$ in one system will induce spikes at times
$t+\tau^C$ in the other system (see light blue arrows) and $t+\tau^K$ in the first system (black arrows) by
mutual coupling and self-feedback, respectively. The spike in the second system returns after a round-trip time
$2\tau^C$. Thus, we have excitation events at times $t+\tau^K$ and $t+2\tau^C$ in each subsystem. The spikes induced by these two sources
of excitation become coherent, i.e., in resonance, when the delay time due to a round trip to the other system and back
again, i.e., $2\tau^C$ matches with the self-coupling delay $\tau^K$. The same argument holds for integer multiples of $\tau^K$ and $2\tau^C$, leading 
to the following condition
\begin{align} \label{eq:relspike}
 N^K \tau^K=N^C 2\tau^C,
\end{align}
with integer numbers $N^K$ and $N^C$. 

Using this notation, Fig.~\ref{fig:samplespikes} displays some combinations of $N^K$ and $N^C$ for coherently spiking
states. Panels (a) and (b) refer to time delays $\tau_K=3=\tau_C$ with $N^C=1$ and $N^K=2$.
While panel (a) shows subthreshold oscillations after $\tau^K$ and superthreshold
oscillations only after $2\tau^C$, panel (b) corresponds to a resonance yielding only fully pronounced, regular oscillations.
Other values of the self-feedback delay, e.g., $\tau_K=2$ and $\tau_K=4$, result in different
combinations of $N^K$ and $N^C$. See, for instance, panels (c) and (d) that correspond to $N^K=3$, $N^C=1$, and $N^K=3$,
$N^C=2$, respectively.
For superthreshold oscillations, Eq.~(\ref{eq:relspike}) yields the following condition for coherent spiking:
\begin{align}
\label{eq:delayK}
 \tau^K=\frac{2\tau^C N^C}{N^K}.
\end{align}
The corresponding ISIs are given by
\begin{align}
\label{eq:Period}
 T=\frac{2\tau^C}{N^K}=\frac{\tau^K}{N^C}
\end{align}
with minimal integer numbers $N^K$, $N^C$, i.e., the fraction $N^K/N^C$ is
irreducible.

\begin{figure}[th!]
  \centering
  \psfig{file=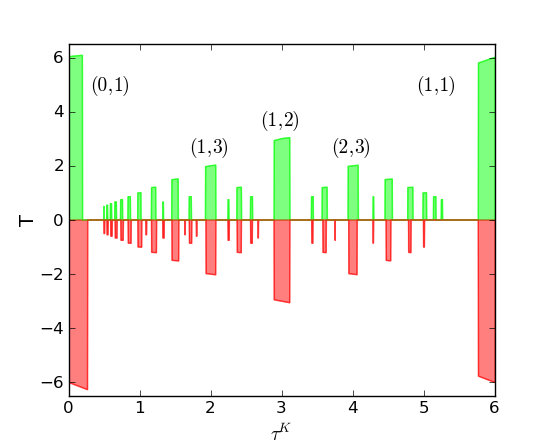,width=3in}
  \caption{Period of possible ISIs of coherent spiking in dependence on $\tau_K$ for $K=0.5$ and $C=0.5$.
The mutual time delay is fixed at $\tau^C = 3$. Green: numerical simulation; red  inverted values: analytical
calculation using Eqs.~(\ref{eq:delayK}), (\ref{eq:Period}) and Eqs.~(\ref{eq:delta_tauK}), (\ref{eq:delta_tauK2}). The parentheses refer to some exemplary values $(N^C,N^K)$. Other parameters as in Fig.~\ref{fig:samplespikes}.
}
  \label{fig:isi_tk}
\end{figure}

Equation~(\ref{eq:Period}) reflects the resonance condition (\ref{eq:relspike}): $N^K$ spikes are induced by self-coupling 
during the round trip time $2\tau^C$ of the mutual coupling. See Figs.~\ref{fig:samplespikes}(b)-(d). 
Figure~\ref{fig:isi_tk} compares
this analytical result to the numerical simulation. The green bars (positive $T$) show the
numerically simulated ISI in dependence on $\tau^K$ for standard deviations smaller than 0.01. The other parameters are
fixed at $K=0.5$, $C=0.5$, and $\tau^C=3$. For larger standard deviations the spiking is not coherent anymore. The red
bars (negative $T$, i.e., inverted to facilitate comparison) show the analytically calculated delay times $\tau^K$ for
which spiking occurs, and the value of the associated
ISI using Eqs.~(\ref{eq:delayK}), (\ref{eq:Period}) and (\ref{eq:delta_tauK}), (\ref{eq:delta_tauK2}). The latter
describe the width of the resonance bars and will be derived later. It can be seen that the analytic results are
in good agreement with the time delay $\tau^K$, for which coherent spiking is found with the corresponding ISI $T$ in 
the numerical simulation. Due
to finite numerical accuracy, not all analytically possible ISI are detected in the time series. For large integers
$N^C$ and $N^K$, coherent spiking does not occur.

It is possible to derive a condition concerning a possible phase shift of spikes in the activator variables $x_1$ and
$x_2$ in the regime of coherent spiking as displayed, for instance, by Figs.~\ref{fig:samplespikes}(c) and (d). If we
find integers $\widetilde{N}^K$ and $N^C$ with
\begin{align}
 \widetilde{N}^K\tau^K=N^C\tau^C,
\end{align}
spikes in the first and second oscillator coincide. 
Thus for $\widetilde{N}^K=N^K/2\in \mathbb N$ leading to $N^K$ even,
there is no phase difference, i.e., we observe in-phase oscillations. Otherwise, for odd $N^K$, the phase shift is $\pi$
corresponding to anti-phase oscillations. Using Eq.~(\ref{eq:delayK}) we are able to predict for which values of $\tau^K$ and $\tau^C$ anti-phase oscillations occur.

\begin{figure}[th!]
  \centering
  \psfig{file=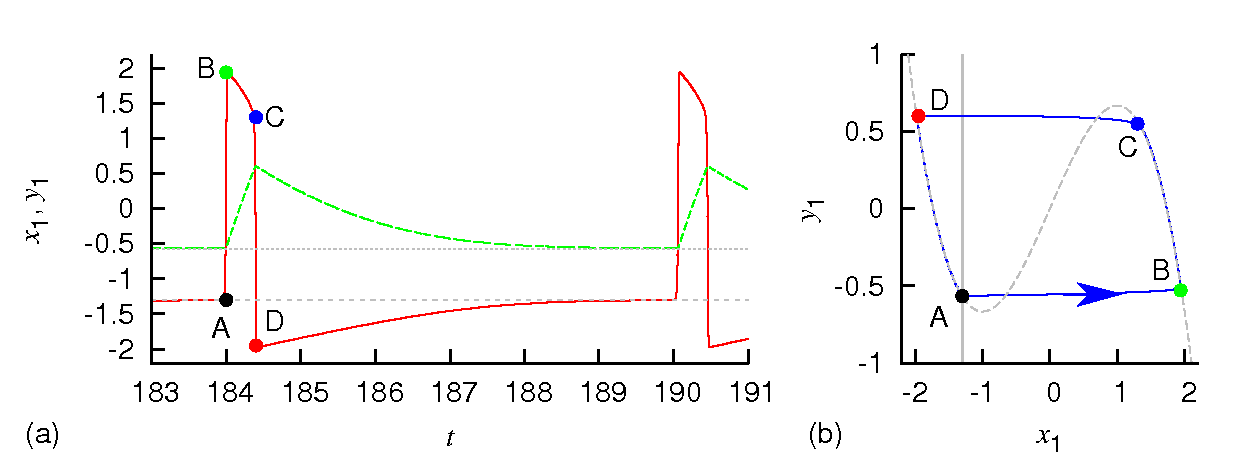}
  \caption{(a) Time series and (b) phase portrait of a numerical solution of
Eq.~(\ref{eq:fhn2_dl_fdb_sym}) for $K=0$, $C=0.5$, and $\tau^C=3$. Other parameters as in
Fig.~\ref{fig:samplespikes}.}
  \label{fig:phaseplane}
\end{figure}

In order to introduce some helpful notation for the following derivation, Figs.~\ref{fig:phaseplane}(a)
and \ref{fig:phaseplane}(b) display a time series and the respective phase portrait of a
typical behavior of the
neural system under consideration. A full excursion in phase space consists of a round trip from A through points B, C,
and D back to A. The times for the transitions from A to B and from C to D are negligible, since the activator variables
$x_1$ change much faster than the inhibitors $y_1$, due to timescale separation $\varepsilon\ll 1$. The transition from B to
C happens close to the right slow branch
of the cubic nullcline during a fixed firing time $T_f=T_{B\rightarrow C}$. In an
earlier publication \cite{SCH08} we derived the following analytical approximation for the firing time $T_f$
\begin{subequations}
\begin{eqnarray}
 T_f&=&\int_B^C\frac{dx_1}{\dot{x}_1} = \int_B^C\frac{1-x_1^2}{x_1+a} dx_1\\
 &=& \left(a^2-1\right)\ln\frac{a+2}{a+1}-a+\frac{3}{2}
 \label{eq:T_B-C}
\end{eqnarray}
\end{subequations}
with $x_1=2$ and $x_1=1$ as an approximation for points B and C, respectively. Thus, we obtain a value of $T_f\approx0.45$
for $a=1.3$. Note that
this approximation is valid for parameter values of $a$ close to $1$. For an improved estimate of $T_f$ further away
from the bifurcation point, one can use the position of the fixed point, i.e., intersection of the nullclines. For
details see Appendix~\ref{app1}.  For the final transition from D to A there remains the time $T_{D\rightarrow
A}=T-T_f$, which will be considered later in this Section.

In Fig.~\ref{fig:isi_tk} one sees that each area of coherent spiking has a certain width. In the following we will
derive an expression for the width $\Delta \tau^K$ of these coherence tongues by using the quantity $T_f$.
For this, we will soften the condition (\ref{eq:relspike}) to within a certain tolerance: If the time shift $|N^K\tau^K-N^C2\tau^C|$ is 
smaller than half the firing time $T_f/2$, spiking still occurs even though Eq.~(\ref{eq:relspike}) is only approximately fulfilled. 
This leads to the following relation
\begin{align}
\label{eq:extension}
 \left|N^K\left(\tau^K\pm\frac{\Delta \tau^K}{2}\right)-N^C2\tau^C\right|\leq \frac{T_f}{2}.
\end{align}
Here $\Delta \tau^K$, i.e., the width of the coherence tongues, acts as tolerance in the timing of the incoming
excitations. Equation~(\ref{eq:extension}) yields an
upper bound for $\Delta\tau^K$
\begin{eqnarray}\label{eq:delta_tauK}
 \left|\Delta\tau^K\right| \leq \frac{T_f+2\left|N^C 2\tau^C-N^K\tau^K \right|}{N^K}.
\end{eqnarray}
Recalling condition~(\ref{eq:relspike}), this simplifies to
\begin{eqnarray}\label{eq:delta_tauK2}
 \left|\Delta\tau^K\right| \leq \frac{T_f}{N^K}.
\end{eqnarray}

\begin{figure}[th!]
  \centering
  \psfig{file=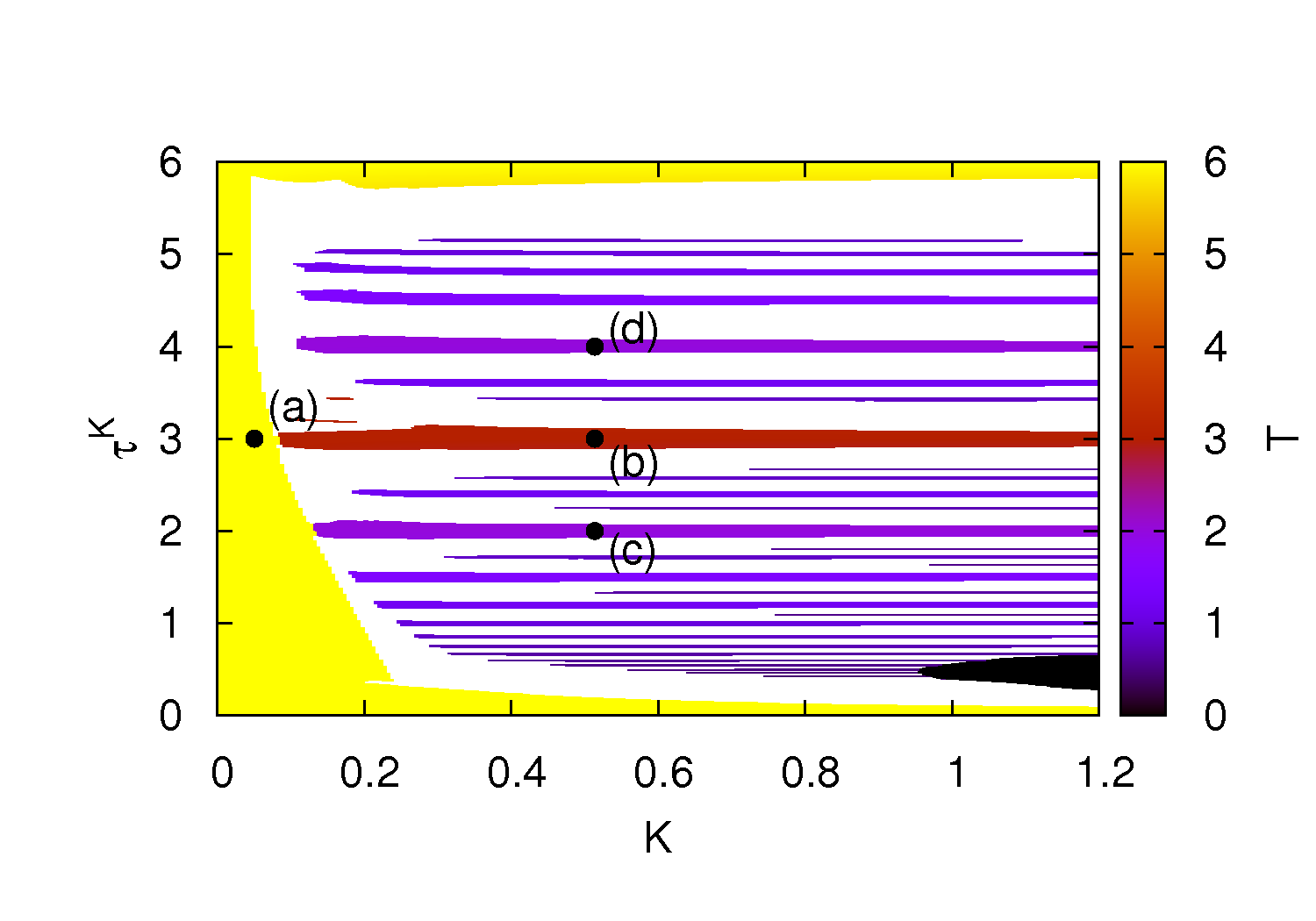}
  \caption{Interspike intervals of the spiking state obtained from numerical simulations in dependence on
$K$ and $\tau^K$ with fixed $\tau^C = 3$ and $C=0.5$. The values marked by (a) to (d) refer to time series displayed in
Fig.~\ref{fig:samplespikes}. Other parameters as in Fig.~\ref{fig:samplespikes}. A spike is considered as an excursion in phase space with $x_1>0$.}
  \label{fig:isi_plane_num}
\end{figure}

For a better analysis of the behavior of the compound system, it is helpful to investigate the ISI.
Figure~\ref{fig:isi_plane_num} shows the ISI as color code in the $\left(K,\tau^K\right)$-plane for fixed
$\tau^C=3$. Note that only spiking with an ISI standard deviation smaller than $0.01$ is depicted. In the white region
the standard deviation of the ISIs is large. Thus, white color marks the region, where no coherent spiking occurs. The horizontal
lines correspond to delays $\tau^K$ in the self-coupling which are in resonance to $\tau^C$. Some combinations of $K$
and $\tau^K$ are marked by black dots (a) - (d) referring to the time series in Fig.~\ref{fig:samplespikes}.

The bright yellow region at small $K$ and $\tau^K$ values refers to a regime, where oscillations with $T=2\tau^C=6$
dominate the dynamics due to mutual coupling, while the self-coupling leads only to subthreshold oscillations. In these cases, 
the self-coupling is too weak or too fast, i.e., small $K$ or small $\tau^K$, respectively, to initiate
additional spikes. Compare also Fig.~\ref{fig:samplespikes}(a). The black region for large $K$ and small $\tau^K$
corresponds to oscillation death due to the refractory phase of the neural oscillator \cite{SCH08}. There the subsystem is not
susceptible to an incoming activating signal.

\begin{figure}[th!]
  \centering
  \psfig{file=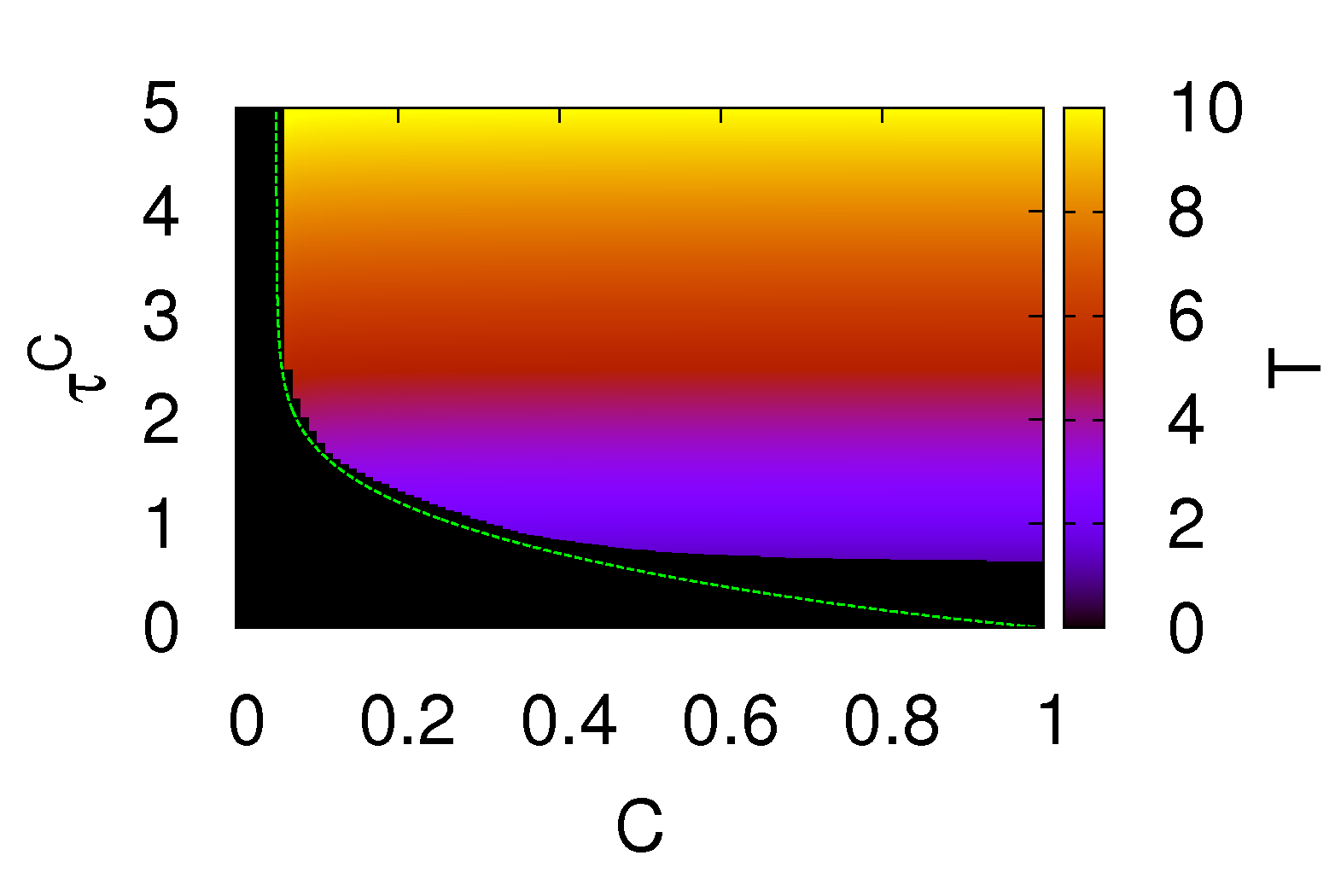,width=4in}
  \caption{Interspike intervals $T$ of spikes in dependence on $C$ and $\tau^C$ for vanishing
self-coupling ($K=0$). The dashed green curve refers to an analytical approximation of the excitation
threshold for $x_2(t-\tau^C)=1$ in Eq.~(\ref{eq:y-min}). Other parameters as in
Fig.~\ref{fig:samplespikes}.
}
  \label{fig:excitation_threshold_a1.3}
\end{figure}

One can also calculate analytically the threshold at which the coherent spiking with $T=2\tau^C$ ceases, i.e., the border between the bright yellow and the white regime in Fig.~\ref{fig:isi_plane_num}. At this boundary, self-coupling becomes strong enough to excite 
superthreshold spikes. To calculate an analogous boundary in Fig.~\ref{fig:excitation_threshold_a1.3}, we set $K=0$ and vary $C$ and $\tau^C$ and extend
the analytic result later to nonzero self-coupling strength $K$ to calculate the boundary in Fig.~\ref{fig:isi_plane_num}. Using the notation introduced
above, the transition from D to A in Fig.~\ref{fig:phaseplane} completes a full excursion. This last transition happens
during a time interval $T_{D\rightarrow A}=T-T_f$. At this stage of the spike, the neural system is susceptible for the
next excitation. For long periods $T$, the system relaxes to the fixed point $x_{\text{FP}}$. For periods $T$ used in 
this paper, however, the next excitation happens already at an earlier point $\text{A}=(x_{1,end},y_{1,end})$. An
analytical estimate for $T_f=T_{B\rightarrow C} $ is given above, see Eq.~(\ref{eq:T_B-C}). 
We can derive a similar
formula for  $T_{D\rightarrow A}$
\begin{eqnarray}\label{eq:T_D-A}
 T_{D\rightarrow A} &=& \int_{D}^{x_{1,end}}\frac{1-x_1^2}{x_1+a} dx_1.
\end{eqnarray}
If the system exhibits spikes with a period $T=2\tau^C$, we have $T_{D\rightarrow A}=2\tau^C-T_f$. Using this
value, Eq.~(\ref{eq:T_D-A}) yields an implicit expression for
$x_{1,end}$
\begin{eqnarray}\label{eq:x_1end}
 x_{1,end} &=& (a-2)\exp\left(\frac{p-2\tau^C+T_f}{a^2-1}\right)-a
\end{eqnarray}
with the abbreviation $p=a(2+x_{1,end})+2-x_{1,end}^2/2$ that depends upon $x_{1,end}$.
Since the relaxation from D to A follows closely the cubic $y$-nullcline, we can calculate a value
for $y_{1,end}$ as follows:
\begin{eqnarray}\label{eq:y_1end}
 y_{1,end} &=& x_{1,end}- \frac{x_{1,end}^3}{3}.
\end{eqnarray}

The mutual coupling ($K=0$, $C\neq 0$) serves as an input in Eqs.~(\ref{eq:fhn2_dl_fdb_sym_x1}) and
(\ref{eq:fhn2_dl_fdb_sym_x2}). Thus, it leads to a temporary, vertical shift of the cubic $y$-nullcline. If the
minimum $(x_{1,min},y_{1,min})$ of this dynamic $y$-nullcline, which includes input from the coupled system, is shifted
beyond the point A$=(x_{1,end},y_{1,end})$, a spike is triggered. The equation for the dynamic $y$-nullcline (for $K=0$) is given by
\begin{align}
\label{eq:y-end1}
 y_1=x_1+\frac{x_1^3}{3}+C\left[(x_2(t-\tau^C)-x_1\right],
\end{align}
which depends on the delayed activator variable $x_2(t-\tau^C)$ that eventually induces the next spike. The
minimum of this nullcline can easily be calculated as
\begin{subequations}
\label{eq:xy-min}
\begin{align}
x_{1,min}&=-\sqrt{1-C}\\
y_{1,min}&=x_{1,min}-\frac{x_{1,min}^3}{3}+C\left[x_2(t-\tau^C)-x_{1,min}\right].\label{eq:y-min}
\end{align}
\end{subequations}
Finally, the condition $y_{1,end}=y_{1,min}$ determines the boundary between coherent spiking and the quiescent
state. Note that the delayed response $x_2(t-\tau^C)$ in Eq.~(\ref{eq:y-min}) remains to be chosen. 

The analytically calculated excitation threshold is shown in Fig.~\ref{fig:excitation_threshold_a1.3} as a dashed green curve for $x_2(t-\tau^C)=1$. This value is motivated by the assumption that $(x_2,y_2)$ is located at point C
in Fig.~\ref{fig:phaseplane}(b) at time $t-\tau^C$. The ISI is depicted in color code. The black region refers to the quiescent state.  

Following the derivation described above, one can also calculate the excitation threshold for $K\neq 0$ and 
fixed mutual coupling parameters $C$, $\tau^C$ in a similar way. For this, Eq.~(\ref{eq:y-end1}) needs to be extended as
follows:
\begin{align}
 y_1&=x_1+\frac{x_1^3}{3}+C\left[(x_2(t-\tau^C)-x_1\right]\nonumber\\
&-K\left[(x_2(t-\tau^K)-x_1\right],
\label{eq:y-end2}
\end{align}
Similarly Eqs.~(\ref{eq:xy-min}) become
\begin{subequations}
\label{eq:xy-min2}
\begin{align}
x_{1,min}&=-\sqrt{1-C-K}\\
y_{1,min}&=x_{1,min}-\frac{x_{1,min}^3}{3}+C\left[x_2(t-\tau^C)-x_{1,min}\right]\nonumber\\
&+K\left[x_2(t-\tau^K)-x_{1,min} \right ].\label{eq:y-min2}
\end{align}
\end{subequations}
Here one has to choose appropriate values for $x_2(t-\tau^C)$ and $x_2(t-\tau^K)$.

\begin{figure}[th!]
  \centering
  \psfig{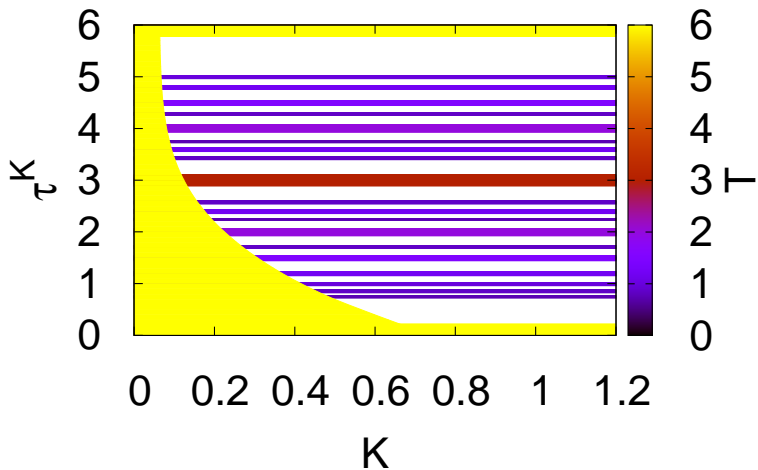}
  \caption{Analytical calculation of the interspike intervals of the spiking state with
    $\tau^C = 3$ and $C=0.5$ using Eqs.~(\ref{eq:delayK}), (\ref{eq:Period}) and (\ref{eq:delta_tauK}),
(\ref{eq:delta_tauK2}). The delayed values $x_2(t-\tau^C)$ and $x_2(t-\tau^K)$ are chosen as $-1.3$ and $2$. Other
parameters as in Fig.~\ref{fig:samplespikes}.
}
  \label{fig:isi_plane_ana}
\end{figure}

Now we assume that the self-coupling delay $\tau^K$ sets the period of the regular spiking leading to $T_{D\rightarrow A}=\tau^K-T_f$. Accordingly, the time $2\tau^C$ in Eq.~(\ref{eq:x_1end}) of $x_{1,min}$ has to be replaced by $\tau^K$. Then the condition derived from $y_{1,end}=y_{1,min}$, i.e., Eqs.~(\ref{eq:y_1end}) and (\ref{eq:y-min2}), and Eqs.~(\ref{eq:Period}) and (\ref{eq:delta_tauK2}) for the ISI lead to
Fig.~\ref{fig:isi_plane_ana} that displays the analytically calculated ISIs. The delayed values $x_2(t-\tau^C)$ and
$x_2(t-\tau^K)$ are chosen as $-1.3$ and $2$, i.e., as points A and B in Fig.~\ref{fig:phaseplane}(b), respectively. This
figure is in good agreement with Fig.~\ref{fig:isi_plane_num}, which shows the numerically simulated ISIs. Not only the
location of the coherence tongues are reproduced by the analytical formulas, but also their widths are in good
agreement.

\section{Nonidentical self-coupling delays}
\label{sec:nonidentical}
As one can see from the previous analysis, the dynamics of system
(\ref{eq:fhn2_dl_fdb_sym}) exhibits a variety of different solutions already for
the simplifying restriction $\tau_1^K = \tau_2^K = \tau^K$. In what follows we
will consider $\tau_1^K \neq \tau_2^K$, in which case one can expect even richer
dynamics. The initial conditions are kept as before, i.e., one subsystem is initialized with a one-time excitation.

\begin{figure}[th]
  \centering
  \psfig{file=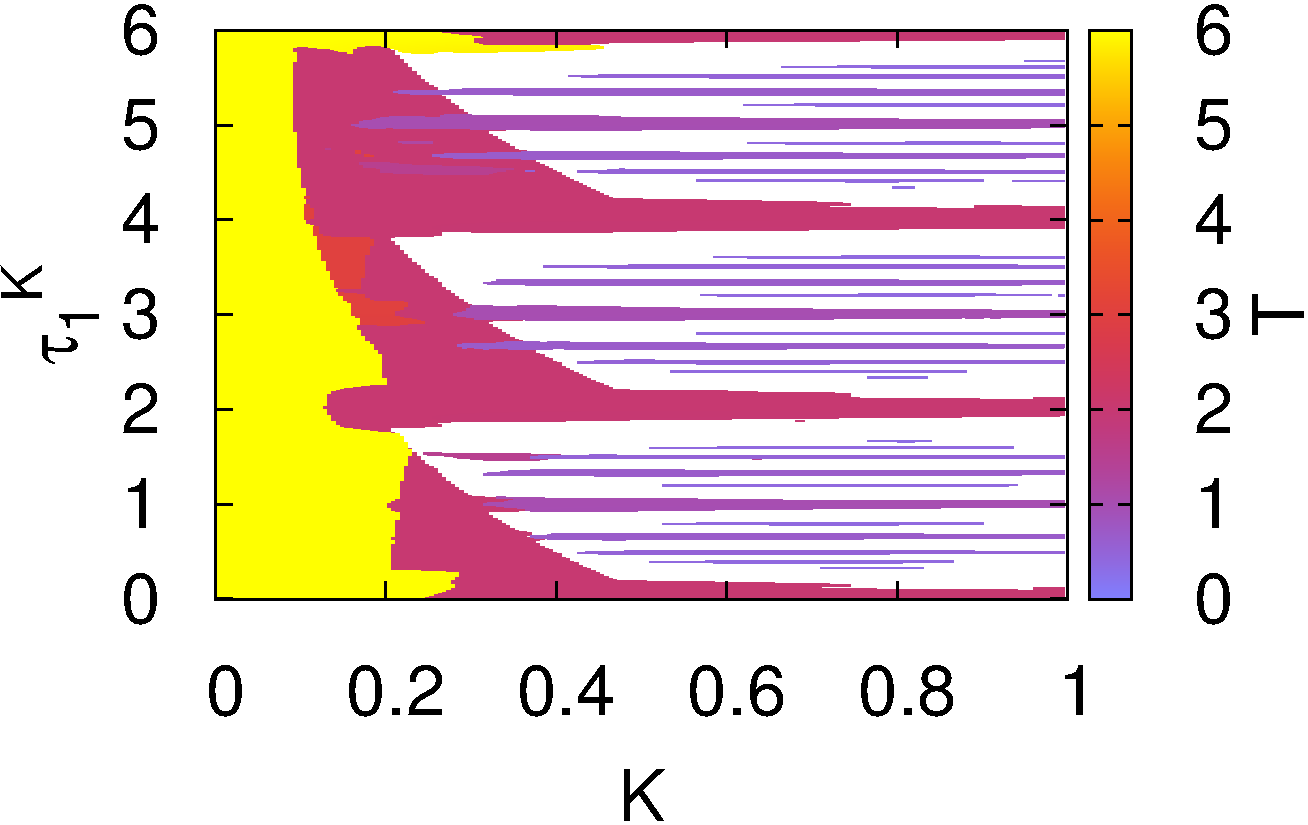,width=4in}
  \caption{Interspike intervals $T$ in the $\left(K,
      \tau^K_1\right)$-plane with $\tau^C = 3, \tau^K_2 = 2$ as color code. Only solutions with an ISI standard deviation smaller than $0.02$ are shown. Other parameters as in Fig.~\ref{fig:samplespikes}.}
  \label{fig:bdinidif}
\end{figure}

In Fig.~\ref{fig:bdinidif}, ISI diagrams in the $\left(K,\tau^K_1\right)$-parameter 
space are plotted for $C = 0.5$, $\tau^C = 3$, and $\tau^K_2 =
2$ in color code. Again, white areas correspond to those
solutions for which the 
ISI standard deviation is larger than a threshold value set to
$0.02$. The colored areas refer to parameters, for which
coherent oscillations similar to a period-1 orbit appear, associated with a
single spike during one period. 

Comparing Fig.~\ref{fig:bdinidif} with
Fig.~\ref{fig:isi_plane_num}, i.e., the case of equal self-feedback delays,
one finds common features like resonances or a boundary of the $T=2\tau^C$ periodic dynamics at small $K$. 
A closer look, however, reveals also distinct differences.
Some of the parameter regions of coherent
spiking appear as tongues at specific ratios of the delays. This time, the regions with $\tau^K_1=\tau_2^K$ and $\tau^K_1=2\tau_2^K$ are more pronounced than for $\tau^K_1=\tau^C$. The reason is that now the fixed self-coupling delay $\tau_2^K$ sets the basic timescale.

Using the same argument as in Sec.~\ref{sec:identical} we can
derive resonance conditions similar to Eq.~(\ref{eq:relspike}). Namely, we
need to require resonance for all pairs of the coupling delays: (i) $\tau^K_1$
and $2\tau^C$, (ii) $\tau^K_2$ and $2\tau^C$, as well as (iii) $\tau^K_1$ and
$\tau^K_2$. The first two assumptions yield, as before,
\begin{equation}
  \label{eq:restau1C}
  N_1^C  2\tau^C = N^K_1  \tau_1^K
\end{equation}
with irreducible integers $N_1^C/N^K_1$, and 
\begin{equation}
  \label{eq:restau2C}
  N_2^C  2\tau^C = N^K_2  \tau_2^K
\end{equation}
with irreducible integers $N_2^C/N^K_2$.
Furthermore, Eq.~(\ref{eq:restau1C}) divided by Eq.~(\ref{eq:restau2C}) immediately
leads to
\begin{equation}
  \label{eq:restau12}
  N_1  \tau_1^K = N_2  \tau_2^K
\end{equation}
with $N_1 = N_2^C  N_1^K/d$ and $N_2 = N_2^C  N_1^K/d$, where $d$ is the greatest common
divisor of $N_2^C  N_1^K$ and $N_2^C  N_1^K$. Therefore, if the relations (\ref{eq:restau1C}) and
(\ref{eq:restau2C}) are satisfied, system (\ref{eq:fhn2_dl_fdb_sym}) performs coherent spiking.  Note that for any three rational numbers $\tau^C$, $\tau^K_1$, and $\tau^K_2$ there always exist corresponding integers $N_1^C$, $N_1^K$, $N_2^C$, $N_2^K$, $N_1$ and $N_2$, for which the Eqs.~(\ref{eq:restau1C})-(\ref{eq:restau12}) hold.

The period of the coherent
solution is also obtained by analogy with the previous case.
First, we find from Eqs.~(\ref{eq:restau1C}) and (\ref{eq:restau2C})
\begin{eqnarray}
  \label{eq:T1_tauK1}
  T_1 &= \dfrac{2\tau^C}{N_1^K} = \dfrac{\tau_1^K}{N_1^C}, \\
  \label{eq:T2_tauK2}
  T_2 &= \dfrac{2\tau^C}{N_2^K} = \dfrac{\tau_2^K}{N_2^C}.
\end{eqnarray}
Finally using Eq.~(\ref{eq:restau12}) we have
\begin{equation}
  \label{eq:T3_tauK12}
  T_3 = \dfrac{\tau_1^K}{N_1} = \dfrac{\tau_2^K}{N_2}.
\end{equation}
Thus, the estimated period is 
\begin{equation}
  \label{eq:3tau_period}
  T = \min\{T_1, T_2, T_3\}.
\end{equation}
As a consequence, if $N_1^K, N_2^K, N^C, N_1$, and $N_2$ are large, the estimated oscillation period becomes small. Then the coherent solution cannot be realized due to the refractory phase of the neural subsystems.

In order to determine the width of the areas for coherent spiking in the $\left(K,
\tau^K_1\right)$-plane, we use a similar reasoning as before. From Eqs.~(\ref{eq:restau1C}) and (\ref{eq:restau12})
we get
\begin{equation}
  \label{eq:dtau1C}
  \left| N_1^K \left( \tau_1^K \pm \frac{\Delta \tau_1^K}{2} \right) -
    N^C 2 \tau^C \right| \leq \frac{T_f}{2}
\end{equation}
and 
\begin{equation}
  \label{eq:dtau12}
  \left| N_1 \left( \tau_1^K \pm \frac{\Delta \tau_1^K}{2}
    \right) - N_2 \tau_2^K \right| \leq \frac{T_f}{2},
\end{equation}
respectively. As a result, the width of the regular spiking regime simplifies to 
\begin{equation}
  \label{eq:dtau1}
  \left| \Delta \tau_1^K \right| \leq \min \left\{ \dfrac{T_f}{N_1^K},
  \dfrac{T_f}{N_1} \right\}.
\end{equation}

\begin{figure}[th]
  \centering
  \psfig{file=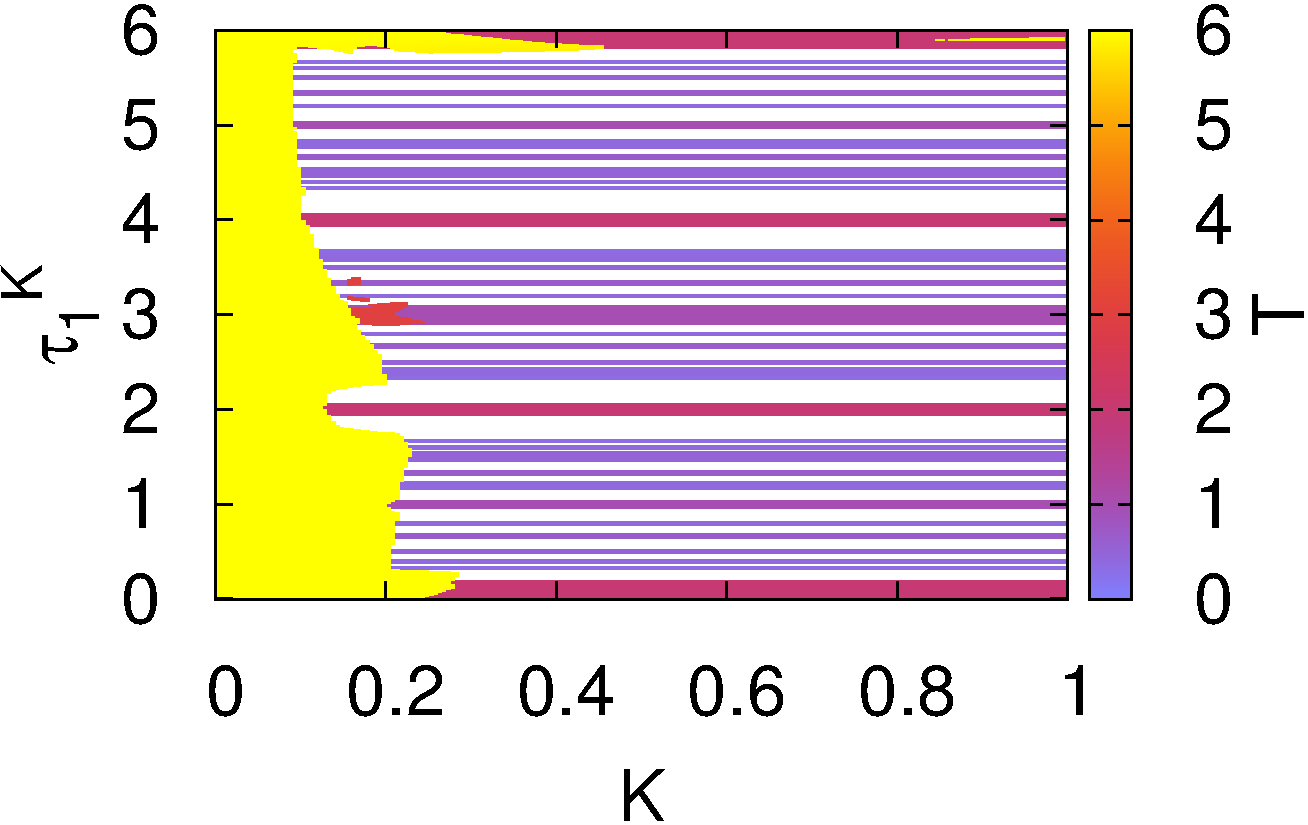,width=4in}
  \caption{Analytical approximation of tongues for
    coherent spiking with period less than $2\tau^C$ in the $\left(K,
      \tau^K_1\right)$-plane ($\tau^C = 3, \tau^K_2 = 2$). The firing
    time $T_f$ is set to $0.38$ (see Appendix~\ref{app1}). The yellow region
    for the solution of period $2\tau^C$ at small $K$ is calculated
    numerically. Other parameters as in Fig.~\ref{fig:samplespikes}.
}
  \label{fig:bd_anal}
\end{figure}

\begin{figure}[th]
  \centering
  \psfig{file=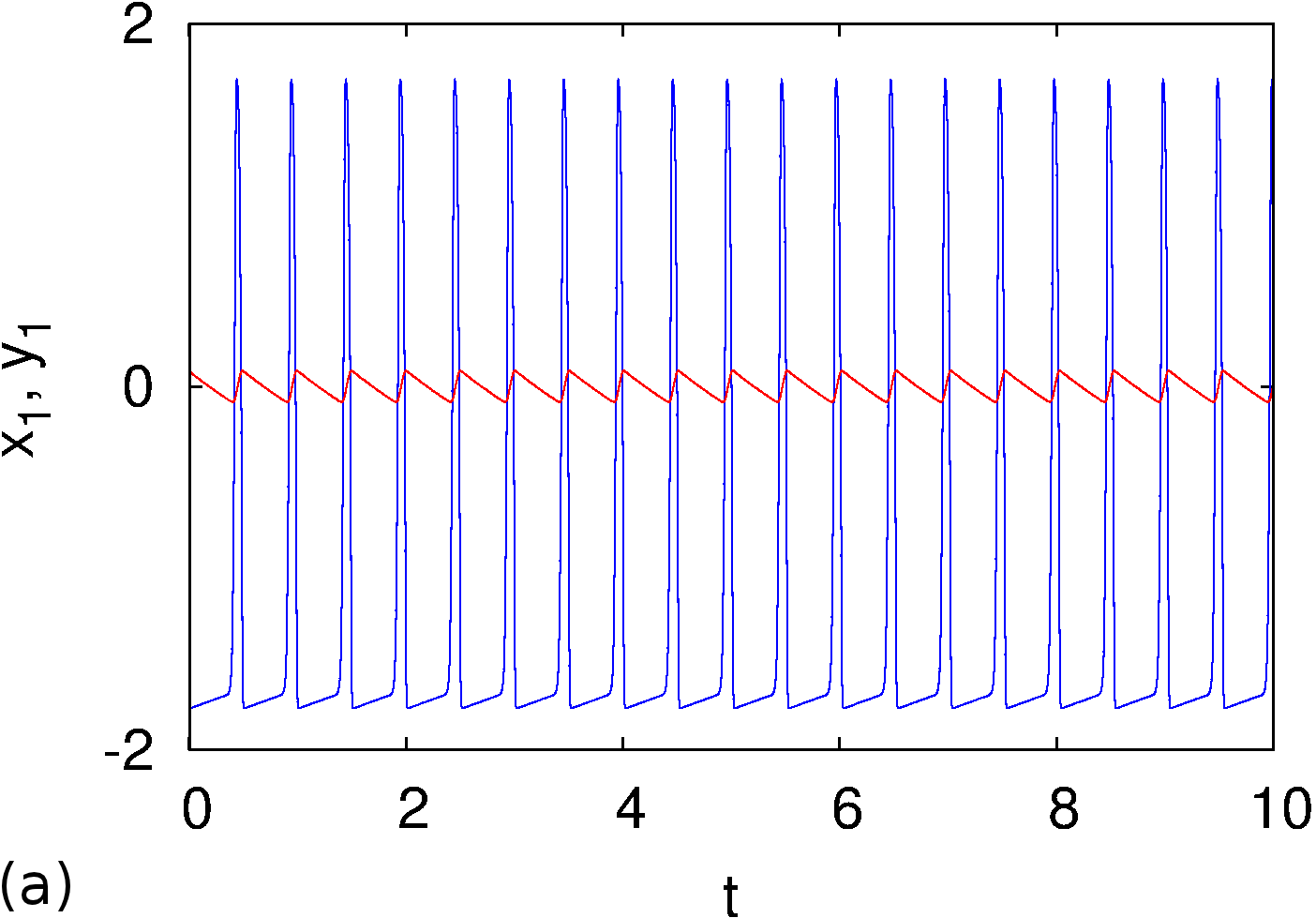,width=0.48\linewidth}
  \psfig{file=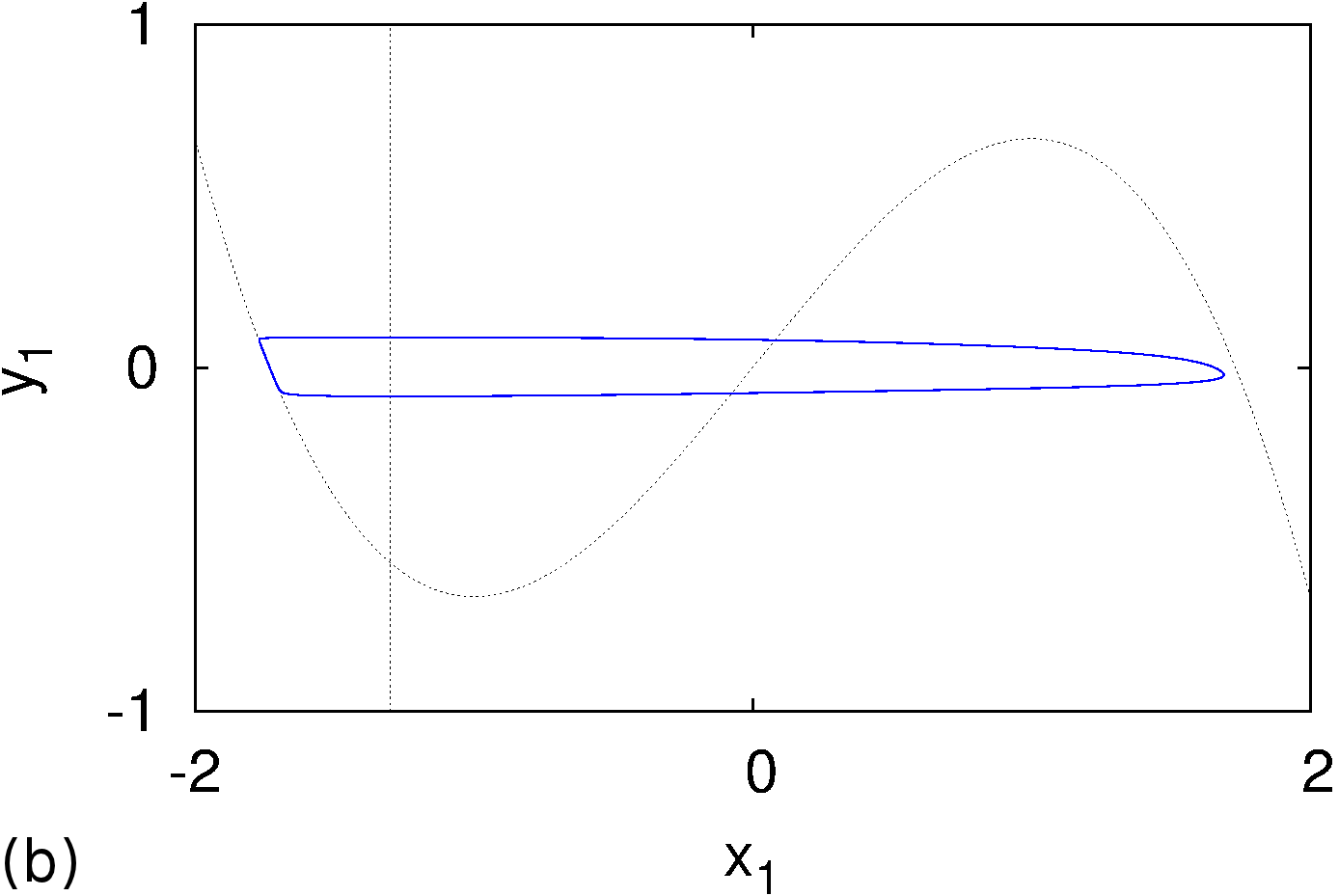,width=0.48\linewidth}
  \caption{Time series and phase portrait of the first
    subsystem variables $x_1$ and $y_1$ as blue and red curves for the 
    periodic solution with period $T \approx 0.5$. Self-coupling parameters: $K = 0.5, \tau_1^K 
    = 0.5, \tau_2^K = 2$. Other parameters as in Fig.~\ref{fig:samplespikes}.}
  \label{fig:solutT05}
\end{figure}

Figure~\ref{fig:bd_anal} shows the analytical estimate for the tongues of
coherent spiking corresponding to Eqs.~(\ref{eq:restau1C}) and
(\ref{eq:restau2C}). The region for the solution of period $T \approx
2\tau^C$ is obtained numerically at small feedback gains $K$.
There is a good correspondence to Fig.~\ref{fig:bdinidif}, however, 
some approximated regions for solutions with small periods are
wider than the simulated ones. The reason is that the approximation for the
firing time $T_f$, given by Eq.~(\ref{eq:T_B-C_new}), is still too rough. For
small periods, the solution looks as exemplarily depicted in Fig.~\ref{fig:solutT05}, and the derivation for $T_f$ does not hold anymore.

As a generalization of our previous analytical approach, we can derive a relation between all three delay
times in general form. Provided that the parameters $C$ and
$K$ are large enough to yield superthreshold excitation, the spike of the first
neuron at time $t$ 
will induce spikes at times $t + \tau_1^K$ and $t +
2\tau^C$. Similarly the second neuron will spike at time $t + \tau^C$, as
well as at times $t +  
\tau^C + \tau_2^K$, $t + \tau^C + 2\tau_2^K$, $t +
\tau^C + 3\tau_2^K,\dots$. 
Thus, the first subsystem will 
get also excitation impulses at times $t + 2\tau^C + \tau_2^K$, $t +
2\tau^C + 2\tau_2^K$, $t + 2\tau^C + 3\tau_2^K$, and so on. From this we get
\begin{equation}
  \label{eq:res3dl1}
  m_1\tau_1^K = l_1(2\tau^C + n_1\tau_2^K),
\end{equation}
where $l_1, m_1, n_1$ are arbitrary positive integers. The same
discussion is applicable for the second 
subsystem, and therefore we can also write down the symmetric
condition
\begin{equation}
  \label{eq:res3dl2}
  n_2\tau_2^K = l_2(2\tau^C + m_2\tau_1^K)
\end{equation}
with positive integers $l_2, m_2, n_2$. 
Adding Eqs.~(\ref{eq:res3dl1}) and (\ref{eq:res3dl2}) we
derive
\begin{equation*}
  2 (l_1 + l_2) \tau^C = (m_1 - l_2 m_2) \tau_1^K + (n_2 - l_1 n_1)
  \tau_2^K.
\end{equation*}
Denoting $\tilde{l} = 2 (l_1 + l_2)/d$, $\tilde{m} = (m_1
  - l_2 m_2)/d$, and $\tilde{n} = (n_2 - l_1 n_1)/d$, where $d$ is the
  greatest common divisor of the three numbers $2 (l_1 + l_2)$, $m_1 -
  l_2 m_2$, and $n_2 - l_1 n_1$, we get
\begin{equation}
  \label{eq:ratio3dlgen}
  \tilde{l}  \tau^C = \tilde{m}  \tau_1^K + \tilde{n} \tau_2^K
\end{equation}
with integers $\tilde{l} > 0$, and $\tilde{m}$, $\tilde{n}$ being of
any sign. Note that Eq.~(\ref{eq:ratio3dlgen}) can  
also be obtained directly from adding or subtracting
Eqs.~(\ref{eq:restau1C}) and (\ref{eq:restau2C}).

At last, by moving all the terms onto one side of the
equation and relabeling,
Eq.~(\ref{eq:ratio3dlgen}) can be rewritten in the general form: 
\begin{equation}
  \label{eq:res3dlfin2}
  l \tau^C + m \tau_1^K + n \tau_2^K = 0,
\end{equation}
where $l, m, n$ are arbitrary integers of 
any sign. 
It should be mentioned that the relation similar to Eq.~(\ref{eq:res3dlfin2}) was obtained in Ref.~\citet{ZIG09} for a two-dimensional time-discrete system with several non-equal delays. 

\begin{figure}[th!]
  \centering
  \psfig{file=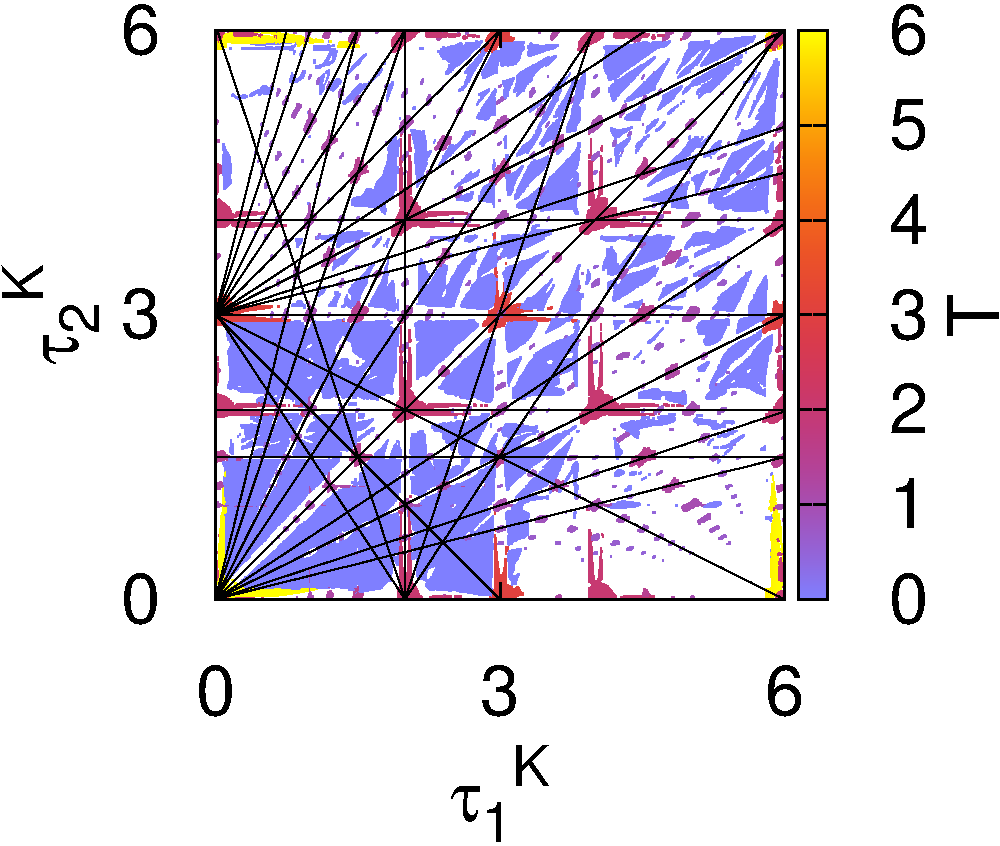,height=3in}
  \psfig{file=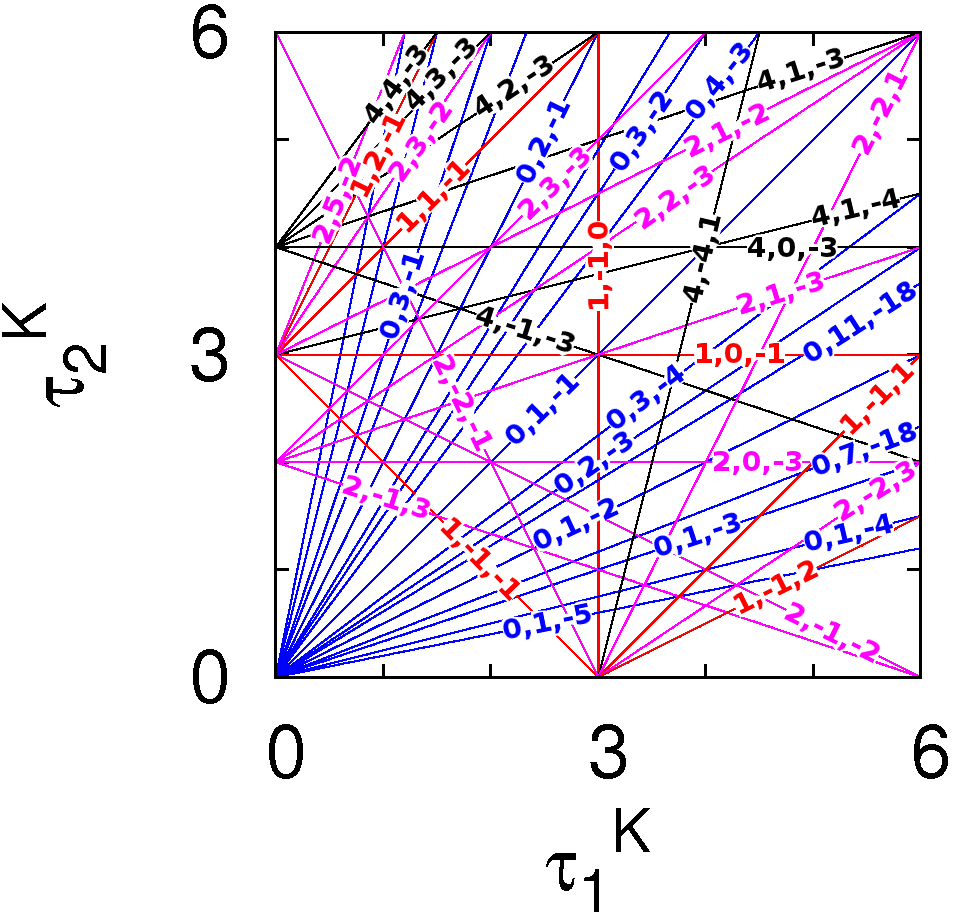,height=3in}
  \caption{(a) Interspike intervals $T$ in the
    $\left(\tau^K_1, \tau^K_2\right)$-plane for $\tau^C = 3, K = 0.5$. The black lines are added according to Eq.~(\ref{eq:res3dlfin2}); (b) resonance lines given by Eq.~(\ref{eq:res3dlfin2}) including the integer values of $l$,
    $m$, and $n$.  Other parameters as in Fig.~\ref{fig:samplespikes}.
  } 
  \label{fig:acflines}
\end{figure}

If we fix $\tau^C$, Eq.~(\ref{eq:res3dlfin2}) defines different lines in the $(\tau_1^K,
\tau_2^K)$-plane, depending on $l, m$, and $n$. Figure~\ref{fig:acflines}(a) visualizes multiple combinations as black lines on top of the ISI $T$ shown in color code. The regions, in which periodic solutions were found
numerically, accumulate along the lines that are given by Eq.~(\ref{eq:res3dlfin2}).
In Fig.~\ref{fig:acflines}(b) these analytically obtained conditions are separately shown including the values of $l$, $m$, and $n$. The analytical results are in excellent agreement with the numerical calculations.

\section{Bursting and autocorrelation function}
\label{sec:bursting}
In Fig.~\ref{fig:acflines}(a) of the previous section, the lines 
obtained analytically from the resonance condition
(\ref{eq:res3dlfin2}) agree very well with the numerical
results on the ISI of regular spiking. However, the  
regions of periodic spiking fill only a part of the
parameter plane. The ISI approach does not allow for an analysis of
bursting-type solutions. There, the time series exhibits bunched spiking patterns and thus different timescales.

Therefore, we consider the autocorrelation function (ACF)
$\Psi(s)$ as an alternative tool for the analysis of the 
coupled system dynamics \cite{HAU06}. The ACF $\Psi(s)$ of an arbitrary time
series $x(t)$ is defined as 
\begin{eqnarray}\label{eq:acf}
  \Psi (s) =  \frac{1}{\sigma^2} \left\langle [x(t-s) - \langle x \rangle]
    [x(t) - \langle x \rangle] \right\rangle, 
\end{eqnarray}
where the averages $\langle \cdot \rangle$ are calculated over the
whole simulated time interval. The value $\sigma$ denotes the standard 
deviation of $x(t)$, i.e., $\sigma^2 = \left\langle [x(t) - \langle x
  \rangle]^2 \right\rangle$.

To determine whether a given solution shows quasi-periodic
neural activity, we calculate the ACF $\Psi(s)$ of
$x_1(t)$, which has several maxima.
The first, trivial maximum is, obviously, obtained at $s =
0$ and equals 1. The second largest maximum marks the
best coincidence between the original and the shifted time series, and is
found for $s = s^*$, which characterizes the length of the repeated
(periodic or quasi-periodic) pattern, and in case of periodic spiking
equals the ISI.

\begin{figure}[t]
  \centering
  \psfig{file=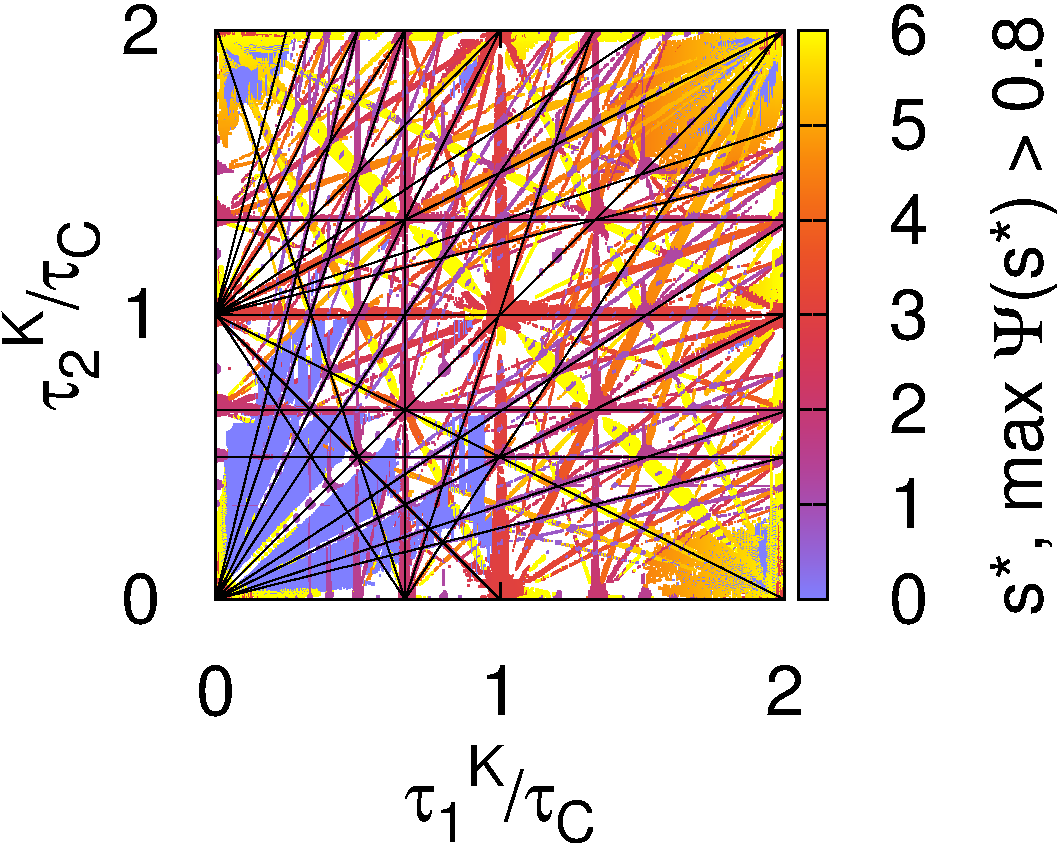,width=0.48\linewidth}
  \caption{Maxima of the autocorrelation function $\Psi$ in the $\left(\tau^K_1, \tau^K_2\right)$-plane for $\tau^C = 3, K = 0.5$. The black lines lines are given by the 
Eq.~(\ref{eq:ratio3dlgen}). Other parameters as in Fig.~\ref{fig:samplespikes}.
}
  \label{fig:reslines}
\end{figure}

The result of this alternative analysis based on the ACF is depicted
in Fig.~\ref{fig:reslines}. In this figure, the resonances of
Eq.~(\ref{eq:res3dlfin2}) (Fig.~\ref{fig:acflines}(b)) are fully
visible and well separated from each other, cf. the yellow and red areas in Fig.~\ref{fig:reslines} that are missing in Fig.~\ref{fig:acflines}(a).

\begin{figure}
  \centering
  \psfig{file=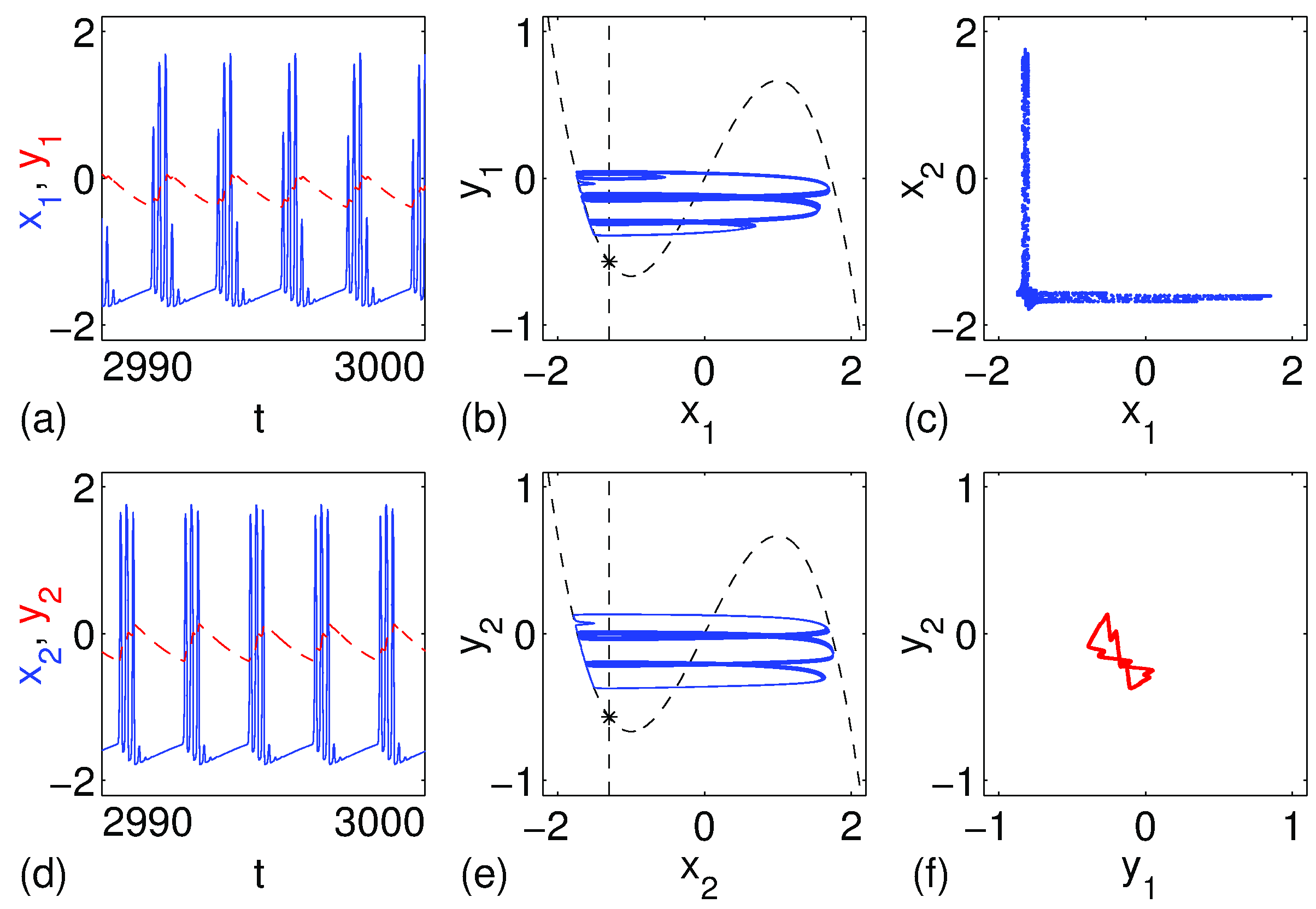,width=0.8\linewidth}
  \caption{Time series (a, d) and phase space projections (b, c, e, f)
    of $x_1, y_1$ and $x_2, y_2$ 
    for the bursting-type solution
    of the system (\ref{eq:fhn2_dl_fdb_sym}) with $K = 0.5, \tau^K_1 =
    2.2, \tau^K_2 = 2$. Other parameters as in Fig.~\ref{fig:samplespikes}.}  
  \label{fig:traj1}
\end{figure}

The ACF allows for investigation of both periodic and quasi-periodic behavior. Figure~\ref{fig:traj1} shows time
series and phase space projections for bursting-type solutions. Since the time series consists of repeated bunches of spikes, the mean interspike interval is no longer a good measure. This is due to two different timescales in the activator variables $x_1$ and $x_2$. To investigate this bursting-type behavior we use again the ACF analysis.

\begin{figure}[th!]
  \centering
  \psfig{file=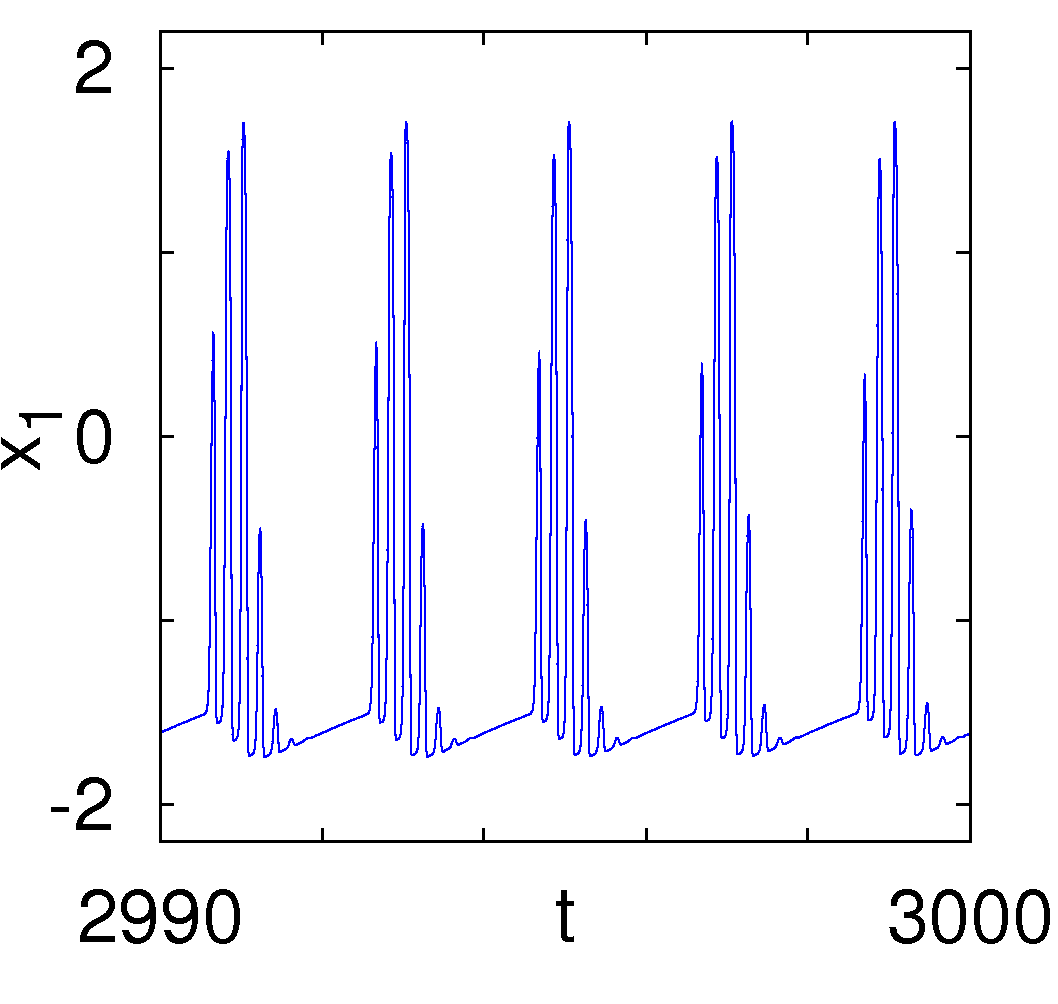,height=2in}
  \psfig{file=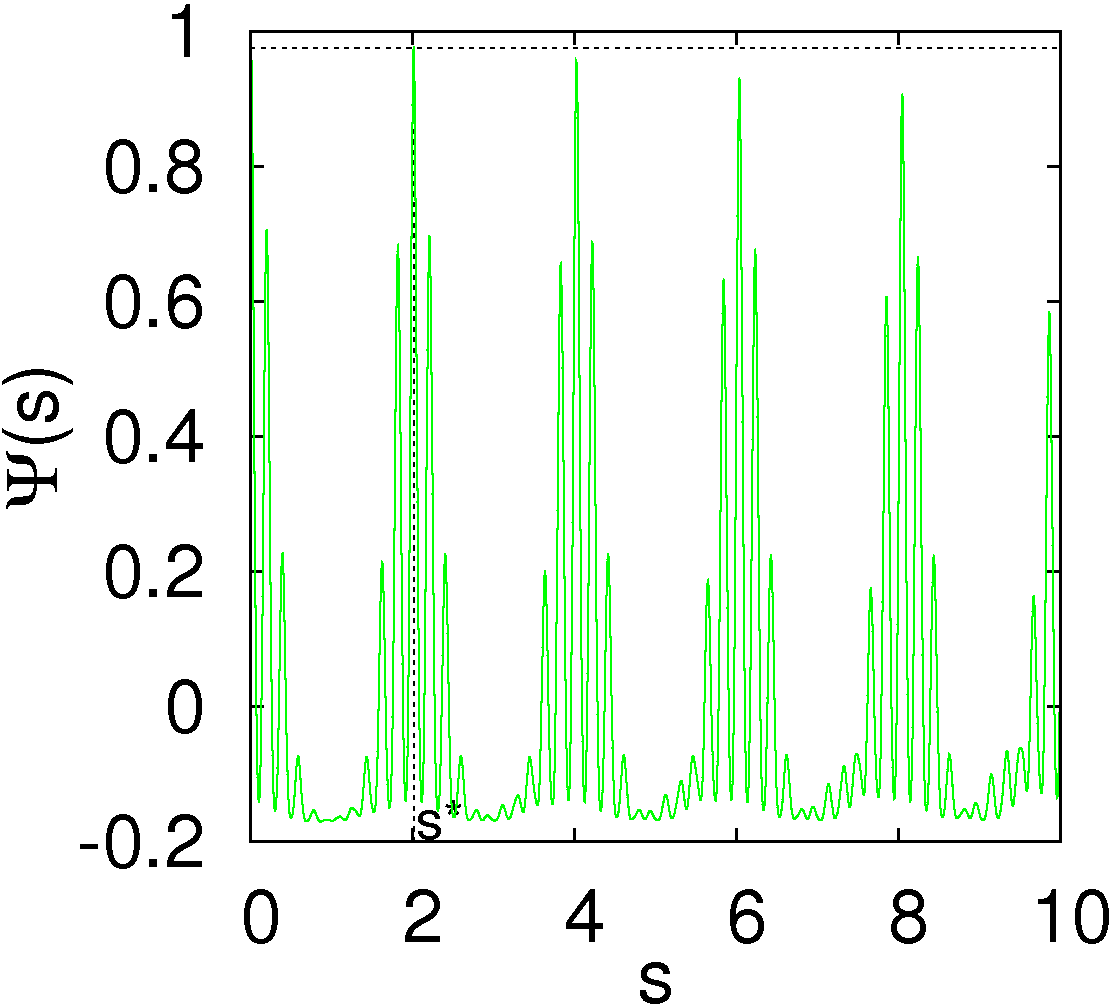,height=2in}
  \caption{(a) Time series of the first activator $x_1$
    and (b) its autocorrelation function $\Psi$, for the same solution as in the
    Fig.~\ref{fig:traj1}. The dotted horizontal line in panel (b) refers to the chosen threshold for correlation detection. Other parameters as in Fig.~\ref{fig:samplespikes}.
  }
  \label{fig:acfx1smp}
\end{figure}

Figure~\ref{fig:acfx1smp}(b) displays the ACF of the
time series from Fig.~\ref{fig:traj1}. In
Fig.~\ref{fig:acfx1smp}(a) the time series for the first activator
$x_1$ is shown again for convenience and Fig.~\ref{fig:acfx1smp}(b) presents the
ACF $\Psi(s)$. It can be seen that the ACF approaches unity for the second time -- after the trivial perfect correlation at $s=0$ -- at a displacement of $s = s^* \approx 2.01$, which equals the repetition period of the bursting pattern. The bursts can also be resolved by the ACF as the fast oscillations. To conclude, the ACF enables to distinguish between inter-burst and intra-burst timescales.

\section{Conclusion}\label{sec:conclusion}
We have investigated effects of heterogeneous time delays for mutual
and self-coupling of a simple network motif that consists of two
neural systems. This setup is realized by two delay-coupled elements
of FitzHugh-Nagumo type, which is a paradigmatic model of neural
interaction. The two subsystems operate in the excitable
regime, and excitation occurs due to the incoming delayed
signals via both mutual and self-coupling.

At first, we have considered identical self-coupling delays and analyzed the regular periodic dynamics on the basis of the mean interspike interval. For small feedback strengths, the system exhibits regular behavior with the period of about twice the mutual coupling delay. With increasing feedback strength, however, the self-feedback term becomes stronger and the system can perform more frequent spikes due to this additional source of excitation. This happens if the mutual coupling and self-feedback delays are in
resonance. In the parameter plane of self-coupling delay and strength, the regions where the regular oscillations exist, resemble stripes emerging from the resonance with the mutual couping delays. We have provided an analytical formula for these resonance conditions and the period of the synchronized oscillations. A comparison with numerical simulations shows excellent agreement.

Next, we have focused on the case of non-identical self-coupling delays. We have considered two-dimensional projections of the parameter space and measured the regularity of the dynamics by the mean interspike interval as well as by the autocorrelation function. Similar to the case of identical self-coupling delays, we observe synchronized periodic dynamics, if the three delay times satisfy a resonance condition. We have also derived a formula for the period of the synchronized regular dynamics, and compared the theoretical results with our numerical simulations.

Finally, we have studied bursts of subsequent excitation spikes. To identify such solutions the measure of interspike intervals is not appropriate any more, and we have used the autocorrelation function as an alternative tool for the analysis of the dynamics.

In conclusion, we have shown that heterogeneous time delays give rise to regular synchronization patterns of different periods, which depend upon resonance conditions of the involved time delays. 

We have restricted our investigations to local dynamics of type-II excitability related to a Hopf bifurcation (FitzHugh-Nagumo model). It is also interesting to consider other models describing, for instance, type-I excitability involving a saddle-node bifurcation on an invariant cycle or physiologically oriented models such as Hodgkin-Huxley- or Morris-Lecar-like equations, but those studies are beyond the scope of the present paper. Another important direction for future research is to increase the number of elements. First approaches to study delayed coupling in large networks have already been reported (see references in Sec.~\ref{sec:intro}). In principle, the results of the presented work can also be applied to coupled systems of more than two elements. Then one has to carefully analyze the combinatorics of the various exciting self-feedback and cross-coupling pluses based on the given network topology. In the case of networks, asymmetries in the coupling strengths can also become important, e.g., if multiple subthreshold excitations accumulate. In addition, different coupling strengths play a crucial role in the presence of both excitatory and inhibitory connections in neural networks. In the presented study of two coupled elements, however, incoming pulses result in all-or-nothing events: if the signal is large enough, it will trigger a full-scale excitation. Otherwise, only a small, subthreshold excitation is possible. 

\nonumsection{Acknowledgments}
\noindent This work was partially supported by DFG in the framework of SFB 910. PH acknowledges support by the BMBF (grant no. 01GQ1001B). We thank Y. Maistrenko, W. Kinzel, I. Kanter, and M. Dahlem for valuable discussions.

\appendix{Approximation of the firing time}
\label{app1}

The formula (\ref{eq:T_B-C}) was derived only for $a$ close to unity, i.e., close to the bifurcation point. For
larger values of $a$, however, a similar estimate holds. In fact, the idea is to integrate in time along the right
branch of the cubic nullcline from the point $B$ to the point $C$ (See Fig.~\ref{fig:phaseplane}). If $a$ is close to 1, the
coordinates of these points can be approximated by $B(2, -2/3)$ and $C(1, 2/3)$. For large $a$ this approximation is rather bad.

To improve the formula one can do the following. The $x$-coordinate of point $A$ cannot exceed $-a$, because the
cycle cannot cross the fixed point $P(-a, a^3/3 - a)$. Therefore, the maximum coordinates for $A$ are
approximately $(-a, a^3/3 - a)$. Since the transition from the left branch to the right branch of the cubic
nullcline happens almost instantaneously, the $y$-coordinate of $B$ could also be approximated as $a^3/3 - a$. This
yields $B=(a/2 + \sqrt{12 - 3a^2}/2, a^3/3 - a)$.

To calculate the coordinates of the point $C$ we proceed as follows.
Since the periodic trajectory in
Fig.~\ref{fig:phaseplane}(b) appears to be symmetric with respect to the origin $(0, 0)$, the $y$-coordinate of $C$ can be obtained as $a - a^3/3$, and hence $C(a, a - a^3/3)$. This leads to the following improved approximation for the firing time $T_f$

\begin{eqnarray}
  \label{eq:T_B-C_new}
  T_f &=& \int_{x_B}^{x_C} \frac{1 - x_1^2}{x_1 + a} dx_1  \\
     &=& \dfrac{x_B^2 - x_C^2}{2} - a(x_B - x_C) + (a^2 - 1) \ln
     \dfrac{x_B + a}{x_C + a} \\
     &=& (a^2 - 1) \ln \dfrac{3a + \sqrt{12 - 3a^2}}{4a} - \dfrac{a}{4}
     \left( a + \sqrt{12 - 3a^2} \right) + \dfrac{3}{2}.
\end{eqnarray}
This yields a value of $T_f \approx 0.38$ for $a = 1.3$.


\end{document}